# Deep neural networks can predict mortality from 12-lead electrocardiogram voltage data

Sushravya Raghunath, PhD[ab], Alvaro E. Ulloa Cerna, MS[ab], Linyuan Jing, PhD[ab], David P. vanMaanen, MS[ab], Joshua Stough, PhD[ac], Dustin N. Hartzel, BS[b], Joseph B. Leader, BA[b], H. Lester Kirchner, PhD[b], Christopher W. Good, DO[d], Aalpen A. Patel, MD[ae], Brian P. Delisle, PhD[f], Amro Alsaid, MBBCh[d], Dominik Beer, DO[d], *Christopher M. Haggerty, PhD[abd], *Brandon K. Fornwalt, MD, PhD[abde]

[a]Department of Imaging Science and Innovation, Geisinger, Danville, PA, USA
[b]Biomedical and Translational Informatics Institute, Geisinger, Danville, PA, USA
[c]Department of Computer Science, Bucknell University, Lewisburg, PA, USA
[d]Heart Institute, Geisinger, Danville, PA, USA
[e]Department of Radiology, Geisinger, Danville, PA, USA
[f]Department of Physiology and Cardiovascular Research Center, University of Kentucky, Lexington, KY, USA
*contributed equally

*The electrocardiogram (ECG) is a widely-used medical test, typically consisting of 12 voltage versus time traces collected from surface recordings over the heart. Here we hypothesize that a deep neural network can predict an important future clinical event (one-year all-cause mortality) from ECG voltage-time traces. We show good performance for predicting one-year mortality with an average AUC of 0.85 from a model cross-validated on 1,775,926 12-lead resting ECGs, that were collected over a 34-year period in a large regional health system. Even within the large subset of ECGs interpreted as "normal" by a physician (n=297,548), the model performance to predict one-year mortality remained high (AUC=0.84), and Cox Proportional Hazard model revealed a hazard ratio of 6.6 (p<0.005) for the two predicted groups (dead vs alive one year after ECG) over a 30-year follow-up period. A blinded survey of three cardiologists suggested that the patterns captured by the model were generally not visually apparent to cardiologists even after being shown 240 paired examples of labeled true positives (dead) and true negatives (alive). In summary, deep learning can add significant prognostic information to the interpretation of 12-lead resting ECGs, even in cases that are interpreted as "normal" by physicians.*

Cardiovascular disease is prevalent, costly, and responsible for a large proportion of morbidity and mortality worldwide[1]. Outcomes for many forms of cardiovascular disease could be improved with better screening and risk prediction, especially with electrocardiograms (ECGs)[2,3]. The 12-lead ECG, typically collected at rest, is one of the most widely used cardiovascular diagnostic tests in the world with standard recommendations for assessment of a wide range of cardiac conditions[4]. Despite its widespread use, it is not well adopted as a prognostic tool[3]. Automated approaches to analyzing ECG data to provide enhanced prognostic capabilities may therefore have tremendous impact on cardiovascular disease outcomes worldwide.

The emergence of large clinically-acquired ECG datasets combined with exponential growth in computational power and improvements in deep neural networks has recently enabled significant advancement in the automated interpretation of ECGs[5–8]. For example, in the ambulatory setting, a deep neural network outperformed cardiologists at diagnosing abnormal heart rhythms in a set of 328 single-lead ECGs (with the model being trained on 91,232 ECGs)[9]. In another study leveraging 100,000 12-lead ECGs in the emergency room setting, a deep learning model was superior to traditional signal processing techniques for identifying acute findings[10]. Deep learning has even shown promise for identifying asymptomatic heart dysfunction using a model trained on 35,970 12-lead ECGs[11,12]. However, an automated method to predict clinically relevant future events, such as short-term mortality, directly from ECGs has not yet been developed. Such a predictive tool could be a valuable asset to aid clinicians in cardiac risk stratification with potentially earlier evaluation and management to reduce mortality risk beyond the current widely-used clinical risk models (for example: Framingham[13], TIMI[14], and GRACE[15] scores), and the specific need for better mortality risk scores has been noted with regard to advanced care planning and palliative care[16–18].

We hypothesized that a deep neural network (DNN) could identify novel features recorded in resting 12-lead ECG voltage-time data and that these features could be used to directly predict an important future clinical event, specifically 1-year mortality. We leveraged nearly 1.8 million ECGs (an order of magnitude larger than previous studies) and DNN to show that this hypothesis holds true. Moreover, we demonstrate that a DNN has higher accuracy to predict 1-year mortality than a model developed using the traditional ECG derived measurements and pattern findings that are routinely performed clinically using ECG. Finally, we showed that the predictive accuracy of a DNN was preserved even in the large subset of ECGs interpreted as normal by physicians, and that three different cardiologists generally could not identify the likely subtle features the model was leveraging to predict survival in this subset.

**Results:**

We extracted all 12-lead ECGs from the electronic records of a large regional US health system. After excluding children (age<18), poor tracings, corrupted data, or patients without at least 1 year of follow-up, there were 1,775,926 ECGs from 397,840 patients available for the study, with 194,845 mortality events within 1-year of the ECG acquisition (Supplementary Figure 1 and Table 1). We trained a DNN (details in methods) to aggregate the spatial and temporal features of the voltage-time signals to predict 1-year survival. We evaluated model performance with a 5-fold cross-validation scheme. The average number of ECGs across five folds in the training, validation and test sets were 1,392,384 (144,214 events), 28,357 (11,662 events) and 355,185 (38,969 events), respectively. The loss function was weighted to compensate for the imbalance between proportion of output labels (alive/dead) during training. ECGs from the same patient were not split between train and test sets.

First, we showed that the area under the receiver operating characteristic curve (AUC) for predicting 1-year all-cause mortality was 0.830 using the ECG voltage-time traces alone and improved to 0.847 when age and sex were added as additional basic demographic features (transparent blue bars Figure 1A). During a 12-lead ECG acquisition, all leads are acquired for a duration of 2.5 seconds and three of those 12-leads (V1, II and V5) are additionally acquired for a duration of 10 seconds.

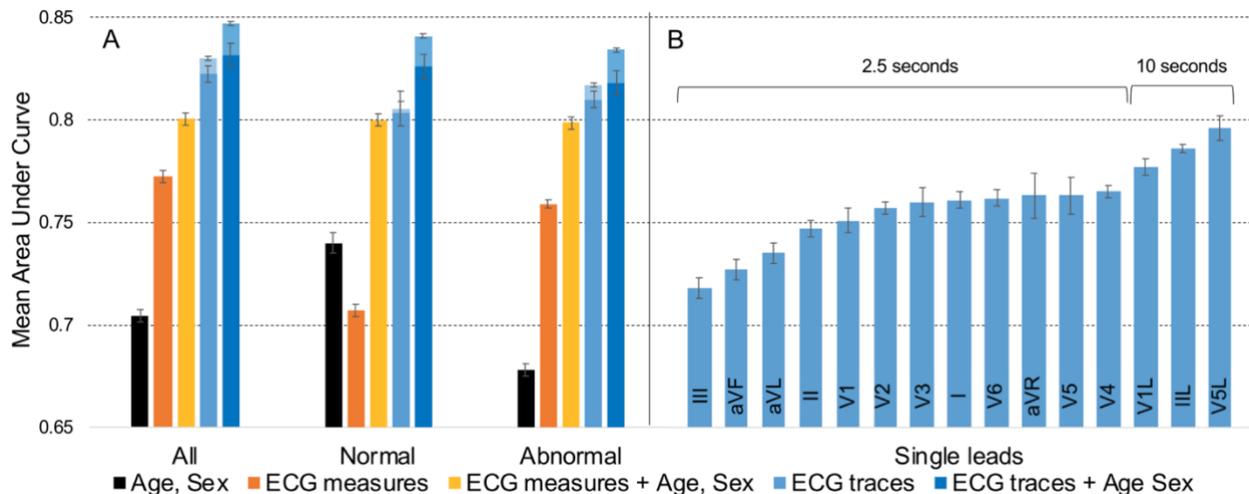

*Figure 1 Summary of model performance to predict one-year mortality (A) For different inputs including (i) age and sex alone (ii) clinically-acquired "ECG measures" (9 numerical values) and 30 diagnostic labels, (iii) ECG measures with age and sex, (iv) ECG voltage-time traces only and (v) ECG voltage-time traces with age and sex. We note that approximately 25% of the ECG traces did not have corresponding measures available in our structured database; hence, we separately report in (iv) and (v) model performance within all available data (transparent bars) and the 75% subset with corresponding structured measures available (solid bars; performance of the models was slightly lower in this subset). Note that models for (i)-(iii) used XGBoost and (iv)-(v) used a DNN. (B) The relative performance of the DNN models using single leads as input (sorted by performance). Area under the receiver operating characteristic curve is reported for the mean of five cross-validation folds. Error bars are standard deviation. (DNN: deep neural network).*

*Table 1 Patient demographics and summary of data distribution across predicted groups (DNN model is trained with all the data, including age and sex, using a likelihood threshold of 0.5).*

| | All | Train (fold=1) | Validation (fold = 1) | Test (fold = 1) | Prediction for test fold = 1 (model trained with *) | |
|---|---|---|---|---|---|---|
| | | | | | Dead | Alive |
| ECGs (N) | 1,775,926 | 1,390,103* | 28,537 | 357,286 | 120,214 | 237,072 |
| Patients (N) | 397,840 | 310,419 | 7,824 | 79,597 | 28,174 | 71,466 |
| Events (N) | 194,845 | 143,327 | 11,565 | 39,953 | 32,540 | 7,413 |
| Age (year) | 63.5 ± 16.4 | 63.5 ± 16.4 | 63.6 ± 16.3 | 63.6 ± 16.4 | 74.1 ± 12.4 | 58.3 ± 15.6 |
| Sex (Male in %) | 50 | 50 | 52 | 50 | 54 | 48 |
| Patterns (30 categorical variables) in % of ECGs | | | | | | |
| Available ECGs (N) | 1,763,405 | 1,380,412 | 28,329 | 354,664 | 119,919 | 234,745 |
| Patients (N) | 396,948 | 309,725 | 7,798 | 79,425 | 28,123 | 71,279 |
| Events (N) | 194,307 | 142,991 | 11,501 | 39,815 | 32,462 | 7,353 |
| Normal | 16.87 | 17.01 | 13.29 | 16.61 | 4.04 | 23.04 |
| Left BBB | 2.82 | 2.80 | 3.51 | 2.84 | 5.34 | 1.56 |
| Right BBBB | 6.20 | 6.20 | 7.10 | 6.13 | 10.22 | 4.04 |
| Incomplete LBBB | 0.41 | 0.41 | 0.58 | 0.42 | 0.86 | 0.20 |
| Incomplete RBBB | 3.19 | 3.19 | 3.31 | 3.19 | 2.90 | 3.34 |
| Atrial fibrillation | 8.60 | 8.53 | 10.94 | 8.68 | 19.59 | 3.11 |
| Atrial flutter | 1.37 | 1.37 | 1.74 | 1.35 | 2.97 | 0.52 |
| Acute MI | 1.07 | 1.05 | 1.58 | 1.09 | 1.47 | 0.90 |
| LVH | 7.69 | 7.72 | 7.12 | 7.63 | 9.16 | 6.85 |
| PVC | 6.59 | 6.52 | 8.10 | 6.74 | 12.50 | 3.79 |
| PAC | 4.89 | 4.89 | 6.21 | 4.81 | 8.01 | 3.17 |
| 1st Degree block | 6.30 | 6.27 | 6.22 | 6.42 | 8.11 | 5.56 |
| 2nd Degree block | 0.15 | 0.15 | 0.19 | 0.14 | 0.24 | 0.10 |
| Fascicular block | 3.12 | 3.12 | 3.45 | 3.11 | 5.23 | 2.03 |
| Sinus Bradycardia | 14.48 | 14.54 | 11.11 | 14.52 | 5.55 | 19.11 |
| Other Bradycardia | 0.15 | 0.14 | 0.16 | 0.16 | 0.20 | 0.15 |
| Sinus Tachycardia | 7.67 | 7.60 | 11.64 | 7.63 | 13.82 | 4.46 |
| VTach | 0.10 | 0.10 | 0.19 | 0.10 | 0.29 | 0.01 |
| SVT | 0.49 | 0.48 | 0.75 | 0.51 | 1.18 | 0.16 |
| Prolonged QT | 5.05 | 4.98 | 6.01 | 5.25 | 8.71 | 3.49 |
| Pacemaker | 4.04 | 4.03 | 4.23 | 4.07 | 7.77 | 2.18 |
| Ischemia | 10.33 | 10.24 | 12.13 | 10.54 | 17.94 | 6.76 |
| Low QRS | 4.52 | 4.44 | 7.06 | 4.60 | 7.30 | 3.21 |
| Intra AV block | 2.13 | 2.12 | 2.82 | 2.14 | 3.99 | 1.20 |
| Prior infarct | 18.20 | 18.11 | 20.75 | 18.32 | 25.18 | 14.82 |
| Nonspecific T wave abnormality | 13.54 | 13.49 | 14.94 | 13.60 | 16.36 | 12.20 |
| Nonspecific ST wave abnormality | 9.22 | 9.18 | 10.74 | 9.26 | 13.04 | 7.33 |
| Left axis deviation | 8.89 | 8.85 | 10.00 | 8.95 | 13.72 | 6.51 |
| Right axis deviation | 1.98 | 1.94 | 3.05 | 2.09 | 3.06 | 1.59 |
| Early repolarization | 0.34 | 0.35 | 0.24 | 0.33 | 0.08 | 0.46 |
| ECG measurements (9 continuous variables) | | | | | | |
| Available ECGs (N) | 1,330,172 | 1,041,859 | 21,103 | 267,210 | 77,111 | 190,099 |
| Patients (N) | 366,643 | 286,405 | 6,891 | 73,347 | 23,614 | 65,599 |
| Events (N) | 125,514 | 91,705 | 8,219 | 25,590 | 19,621 | 5,969 |
| QRS duration (ms) | 96.2 ± 22.7 | 96.2 ± 22.7 | 97.0 ± 239 | 96.3 ± 22.6 | 102.9 ± 29.2 | 93.7 ± 18.7 |
| QT (ms) | 401.4 ± 47.1 | 401.6 ± 46.9 | 396.6 ± 51.1 | 401.8 ± 47.4 | 397.5 ± 59.1 | 403.5 ± 41.5 |
| QTC (ms) | 441.7 ± 35.9 | 441.5 ± 35.7 | 446.7 ± 38.3 | 441.9 ± 36.1 | 461.8 ± 41.4 | 433.8 ± 30.2 |
| PR interval (ms) | 163.2 ± 38.6 | 163.2 ± 38.7 | 161.5 ± 39.0 | 163.4 ± 39.5 | 164.5 ± 51.8 | 163.0 ± 33.1 |
| Vent rate (BPM) | 75.5 ± 18.1 | 75.2 ± 18.0 | 79.5 ± 20.4 | 75.4 ± 18.2 | 84.8 ± 20.8 | 71.6 ± 15.4 |
| Avg RR interval (ms) | 837.7 ± 186.6 | 836.3 ± 186.1 | 798.4 ± 190.9 | 835.7 ± 187.1 | 747.3 ± 182.0 | 871.6 ± 177.0 |
| P Axis | 47.6 ± 26.1 | 47.5 ± 26.0 | 48.2 ± 26.9 | 47.7 ± 26.1 | 49.3 ± 31.4 | 47.0 ± 23.5 |
| R Axis | 22.3 ± 45.9 | 22.3 ± 45.8 | 22.1 ± 48.5 | 22.3 ± 45.9 | 16.7 ± 56.3 | 24.6 ± 40.8 |
| T Axis | 50.3 ± 47.4 | 50.2 ± 47.2 | 54.3 ± 52.3 | 50.7 ± 47.7 | 66.9 ± 63.6 | 44.1 ± 37.9 |
| **N**: number of samples; **ms**: milliseconds; **BPM**: beats per minute; **PVC**: Premature Ventricular Complexes; **PAC**: Premature Atrial Complexes; **VTach**: ventricular Tachycardia; **SVT**: Supraventricular Tachycardia; **MI**: Myocardial Infarction; **BBB**: bundle branch block; **LVH**: Left Ventricular Hypertrophy; **AV**: atrioventricular; **Vent rate**: Ventricular rate | | | | | | |

The model with all 15 ECG voltage-time traces from the 12 standard leads together (12 leads acquired for 2.5 seconds plus 3 leads acquired for 10 seconds) provided the best AUC compared to models derived from each single lead as input (Figure 1B). Models derived from the 10-second tracings had higher AUCs than the models derived from the 2.5-second tracings, demonstrating that a longer duration of data may provide more informative features to the model (Figure 1B). Models trained separately for males and females did not show improved AUC (data not shown).

Next, we showed that the AUC from the DNN model derived from the voltage-time tracings alone was superior to a model that utilized the traditional measurements and patterns/diagnoses which are clinically reported as part of a standard 12-lead ECG findings. We compiled these tabular "ECG measures" from the clinical reports, including 9 continuous numerical measurements (e.g. QRS duration) and 30 categorical ECG patterns (e.g. left bundle branch block) (complete list in Methods). To determine the predictive power of the tabular ECG features we trained an XGBoost (XGB) classifier[19,20] to predict 1-year mortality using these 39 "ECG measures". The performance of the XGB model was 0.772, and this improved to 0.810 with the addition of age and sex (Figure 1A). Note that both of these numbers were significantly (p<0.0001 and p=0.0004, respectively, with paired t-tests) below the AUC of the DNN model derived from the ECG traces either without (0.822) or with (0.831) age and sex as additional variables for the matched data-set (solid blue bars in Figure 1A). The demographics and distribution of ECG measures and patterns for the predicted groups for the DNN model with age and sex is summarized in Table 1, which demonstrates general differences between the predicted groups such as increased heart rate and higher prevalence of certain diagnoses like bundle branch blocks or ectopic beats in the predicted dead group.

Subsequently, we determined the performance of the model in ECGs interpreted as "normal" by the physician compared to those with at least one diagnostic abnormality (which we refer to as "abnormal"). Note that an ECG interpreted as "normal" does not necessarily imply that the ECG was collected from a patient without cardiovascular disease, and we will refer to this as normal ECGs henceforth. On average, there were 59,510 normal and 293,171 abnormal ECGs in each of the 5 test folds. For normal ECGs, the DNN model (trained to predict 1-year mortality from all the ECGs) yielded an AUC of 0.805 and 0.841, respectively, with ECG traces alone and with the addition of age and sex; for abnormal ECGs the model yielded an AUC of 0.817 and 0.834, respectively (Figure 1A). The same overall trend of the DNN utilizing the ECG traces alone being superior to the ECG measures held true across both categories of abnormal and normal ECGs.

To further investigate predictive performance within the overall dataset and the subsets of normal and abnormal ECGs, Kaplan-Meier survival analysis was performed (Figure 2B) using follow-up data available in the EHR for the two groups predicted by the model (alive/dead in 1-year) at the chosen operating point (likelihood threshold = 0.5, sensitivity: 0.76, specificity: 0.77; see Table 2). For normal ECGs, the median survival times (for the mean survival curves of five-folds) of the two groups predicted alive and dead at 1-year were 26 and 8 years, respectively, and for abnormal ECGs, 16 and 6 years, respectively (Figure 2B). A Cox Proportional Hazard regression model was fit for each of the five-folds and mean hazard ratios (with lower and upper bounds of 95% confidence intervals) were: 4.4 [4.0–4.5] in all ECGs, 3.9 [3.6–4.0] in abnormal ECGs and 6.6 [5.8–7.6] in normal ECGs (all p<0.005) comparing those predicted by the DNN to be alive versus dead at 1-year post-ECG. Thus, the hazard ratio was largest in the subset of normal ECGs, and the prediction of 1-year mortality from the DNN was a significant discriminator of long-term survival for up to 30 years after ECG.

Next, we investigated if the features learned by the model are visually apparent to cardiologists and whether they are clinically interpretable. To do this, we chose 401 sets of paired normal ECGs and conducted a blinded survey with three cardiologists. Each pair consisted of a true positive (normal ECG correctly predicted by the model as dead at one year) and a true negative (normal ECG correctly predicted by the model as alive at one year), matched for age and sex. The cardiologists generally had poor accuracy of 55-68% (10-36% above random chance) to correctly identify the normal ECG linked to 1-year mortality. After allowing each cardiologist to study a separate dataset of 240 paired ECGs labeled to show the outcome, their prediction accuracy in repeating the original blinded survey of 401 paired ECGs remained low (50-75% accuracy i.e. 0-50% above random chance) (Figure 3).

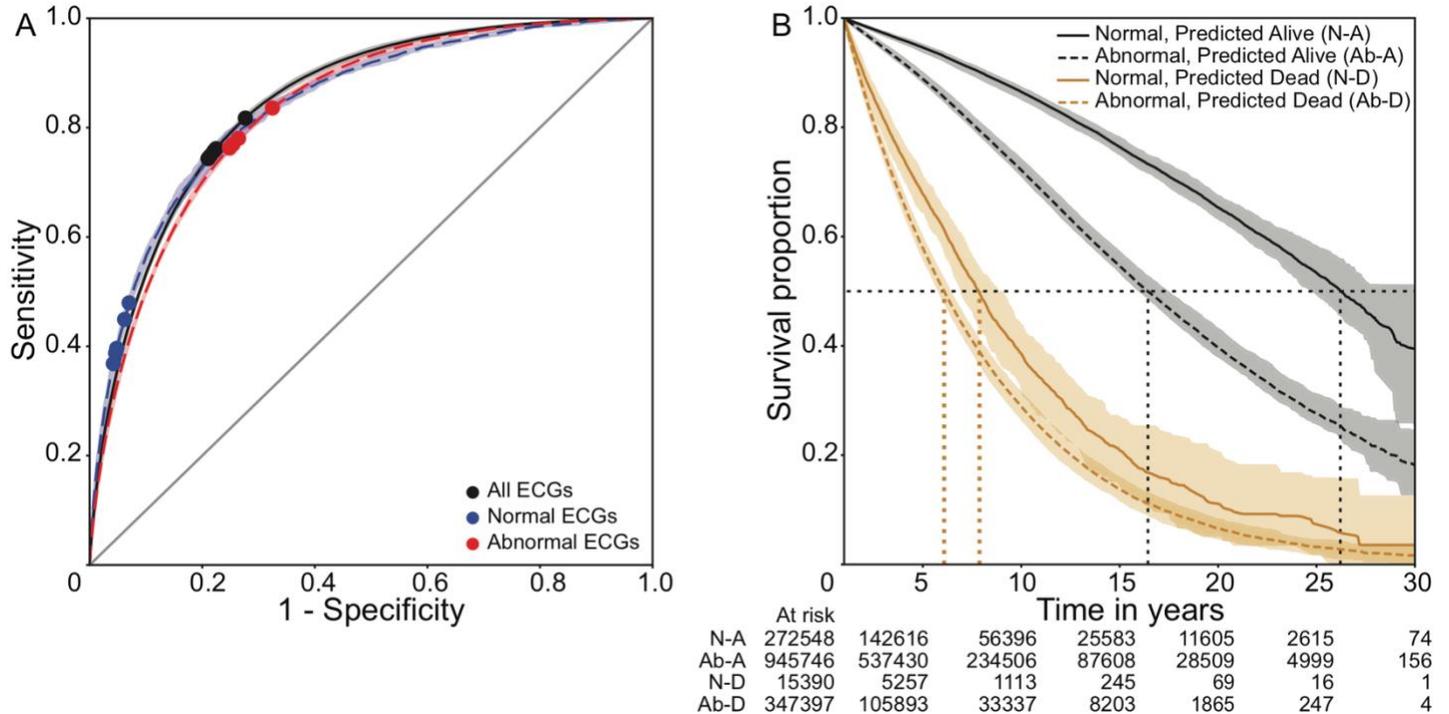

*Figure 2 Receiver operating characteristic (ROC) curves of the trained model for the chosen operating points (threshold on likelihood = 0.5) and corresponding Kaplan-Meier (KM) survival curves. (A) ROC curves (mean of five test folds with shaded 95% confidence region) with operating points for the folds marked for all the data (black), the normal ECG subset (blue) and the abnormal ECG subset (red). (B) The mean KM curves for predicted alive and dead groups in the normal and abnormal ECG subsets of the five folds. The shaded areas are lower and upper bounds of 95% confidence intervals of the five folds. The table shows the at risk population for the given time intervals for all the five test folds combined.*

*Table 2 Summary performance statistics for different operating points of the model trained with all the data. The values are represented as mean ± standard deviation of the five test folds.*

| OP | All data | | | | Abnormal ECGs | | | | Normal ECGs | | | |
|---|---|---|---|---|---|---|---|---|---|---|---|---|
| | Sens | Spec | PPV | NPV | Sens | Spec | PPV | NPV | Sens | Spec | PPV | NPV |
| 0.25 | 0.88 ± 0.01 | 0.64 ± 0.02 | 0.23 ± 0.01 | 0.98 ± 0.001 | 0.90 ± 0.01 | 0.58 ± 0.02 | 0.24 ± 0.01 | 0.98 ± 0.001 | 0.62 ± 0.03 | 0.87 ± 0.01 | 0.14 ± 0.01 | 0.99 ± 0.001 |
| 0.5 | 0.76 ± 0.02 | 0.77 ± 0.02 | 0.29 ± 0.01 | 0.96 ± 0.002 | 0.79 ± 0.02 | 0.73 ± 0.02 | 0.30 ± 0.01 | 0.96 ± 0.002 | 0.42 ± 0.03 | 0.95 ± 0.01 | 0.21 ± 0.01 | 0.98 ± 0.001 |
| 0.75 | 0.54 ± 0.04 | 0.90 ± 0.01 | 0.39 ± 0.02 | 0.94 ± 0.003 | 0.56 ± 0.04 | 0.88 ± 0.01 | 0.39 ± 0.02 | 0.93 ± 0.003 | 0.18 ± 0.04 | 0.99 ± 0.004 | 0.32 ± 0.02 | 0.97 ± 0.001 |
| **OP**: operating point; **Sens**: Sensitivity; **Spec**: Specificity; **PPV**: positive predictive value; **NPV**: negative predictive value | | | | | | | | | | | | |

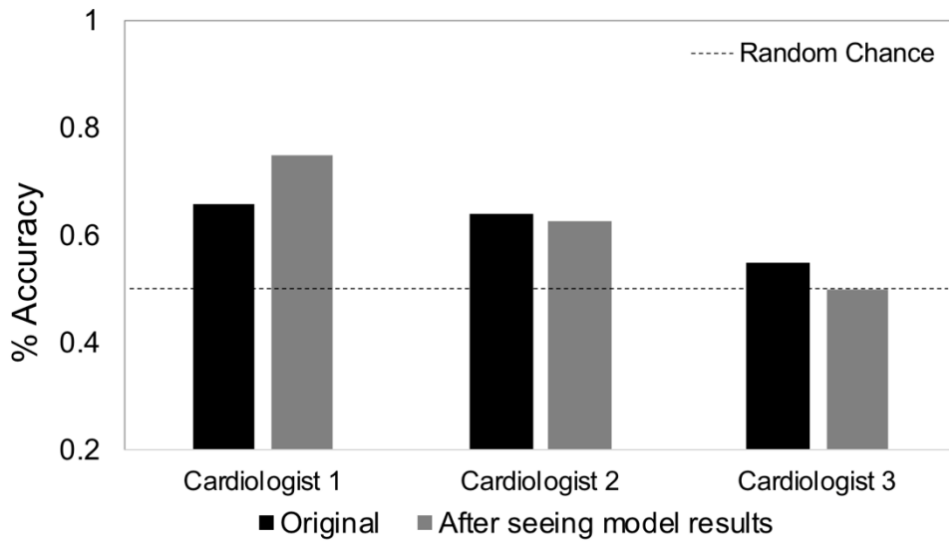

Figure 3 Accuracy for the three cardiologists to correctly identify the true positive ECG (dead within a year) when presented with two 'normal' ECGs corresponding to a paired set of a true positive and true negative (n=401) (black bars). Accuracy is also shown (grey bars) for the same survey after being shown an independent set (n=240) with outcomes labeled. All ECG pairs presented were matched for age and sex.

We chose to predict 1-year, all-cause mortality as an initial target endpoint given its ready availability with high accuracy and less risk of bias related to clinical input, documentation or opinion. Given the good predictive capability we identified using an "all-cause" outcome, we expect that similar models built using voltage-time data will be equally or more accurate for predicting more specific cardiovascular outcomes in future work. Though these data had inherent heterogeneity (which is important for building a generalizable model) since they were collected within a large regional healthcare system with 13 hospitals and hundreds of clinics over almost 4 decades, additional independent datasets will be required to assess more widespread generalizability. Unfortunately, procuring and curating a large external dataset with similar longitudinal outcomes annotation is currently a prohibitive challenge for validation.

**Discussion:**

In summary, we leveraged a large set of nearly 1.8 million ECGs collected from ~400 thousand patients over a period of 34 years to demonstrate the potential for DNNs to automatically predict a highly clinically relevant endpoint (1-year mortality) directly from 12-lead ECG voltage-time data. This potential is evident through several critical findings. First, the DNN model using voltage-time traces outperformed another machine learning model that utilized an extensive collection of 39 clinically-derived ECG features (including both numerical measurements and diagnostic patterns), suggesting that the model is able to identify novel patterns of significant prognostic importance from voltage-time traces. Second, the DNN model not only can predict 1-year mortality with AUC of 0.847, but also shows considerable prognostic value in longer time survival characteristics among the predicted groups with median survival being over 3-times longer in patients with ECGs predicted to be alive versus dead. Finally, despite the canonical wisdom of the high negative predictive value of ECGs[21,22], we found that prediction accuracy remained high even in the large subset of ECGs clinically interpreted as "normal" by a cardiologist. This suggests that features interpreted by the model were not generally apparent to cardiologists upon re-evaluation, even after being shown 240 paired examples of labeled true positives and negatives, further underscoring their novelty. Machine learning therefore has potential to add significant prognostic information to the clinical interpretation of one of the most widely-used medical tests and possibly risk stratify patients who may benefit from preventive interventions.

**Methods:**

*ECG and patient data:*

The Institutional Review Board approved this study with a waiver of consent. We extracted 2.6 million standard 12-lead ECG traces from our institutional clinical MUSE (GE Healthcare, WI) database, acquired between 1984 and 2018. We retained only the resting 12-lead ECGs with voltage-time traces of 2.5 seconds for 12-leads and 10 seconds for 3 leads (V1, II, V5) that did not have significant artifacts and were associated with at-least a year of follow-up or death within a year. This amounted to 1.8 million ECGs where 51% of them were stored at 500 Hz and the remaining were stored at 250 Hz. All data therefore was resampled to 500 Hz by linear interpolation. Characteristics of the final patient population are shown (Table 1). We further cross-referenced the patient identifiers with the most recent clinical encounters and a regularly-updated death index registry at our institution to assign patient status (dead/alive). These data were divided into five folds with five percent of the training data set aside as a validation set. The data were split such that the same patient was not in both train and test sets for cross-validation. The outcomes were approximately balanced in the validation set.

Additionally, the findings within the final ECG clinical reports were parsed to identify 30 diagnostic pattern classes and 9 continuous ECG measurements (all detailed below). Each ECG was defined to be "abnormal" if the pattern label was flagged for at least one diagnostic abnormality. The 9 ECG measurements included *QRS duration, QT, QTC, PR interval, ventricular rate, Average RR interval and P, Q and T-wave axes*. Patterns included *normal, left bundle branch block, incomplete left bundle branch block, right bundle branch block, incomplete right bundle branch block, atrial fibrillation, atrial flutter, acute myocardial infarction, left ventricular hypertrophy, premature ventricular contractions, premature atrial contractions, first degree block, second degree block, fascicular block, sinus bradycardia, other bradycardia, sinus tachycardia, ventricular tachycardia, supraventricular tachycardia, prolonged QT, pacemaker, ischemia, low QRS voltage, intra-atrioventricular block, prior infarct, nonspecific t-wave abnormality, nonspecific ST wave abnormality, left axis deviation, right axis deviation and early repolarization*. The tabular ECG features were only available within 75% of all the available ECGs.

The survival time and patient age were calculated with reference to the date of ECG acquisition and only patients above 18 years of age at the time of ECG were included in this study. Sex was also extracted from the EHR data. Note that death data are highly accurate in our EHR due to regular checks against death index databases, however, survival was not assumed without a known living encounter.

*Model development and evaluation:*

We designed a convolutional neural network (model architecture illustrated in Figure 4) with 5 branches with the input of 3 leads as channels that are concurrent in time, i.e., (Branch 1: [I, II, III]; 2: [aVR, aVL, aVF]; 3: [V1, V2, V3]; 4: [V4, V5, V6] and 5: [V1-long, II-long, V5-long]). Note that each branch represents the 3 leads that were acquired at the same time (during the same heartbeats), for a duration 2.5 seconds. In a typical 12-lead ECG, 4 of these groups of 3 leads are consecutively acquired over a duration of 10 seconds. Concurrently, the "long leads" are recorded over the entire 10-second duration. Thus, the architecture was designed to account for these details since arrhythmias, in particular, cause the traces to change morphology throughout the standard clinical acquisition.

A convolutional block consisted of a 1-dimensional convolution layer followed by batch normalization and rectified linear units (ReLU)[23] activations. The first four branches and last branch consisted of 4 and 6 convolutional blocks, respectively, followed by a Global Average Pooling (GAP)[24] layer. The outputs of all the branches were then concatenated and connected to a series of 6 dense layers of 256 (with dropout), 128 (with dropout), 64, 32, 8 and 1 unit(s) with a sigmoid function as the final layer. We used *Adam*[25] optimizer with a learning rate of $1e^{-5}$ and batch size of 2048 and computed each model branch in parallel on a separate GPU for faster computation.

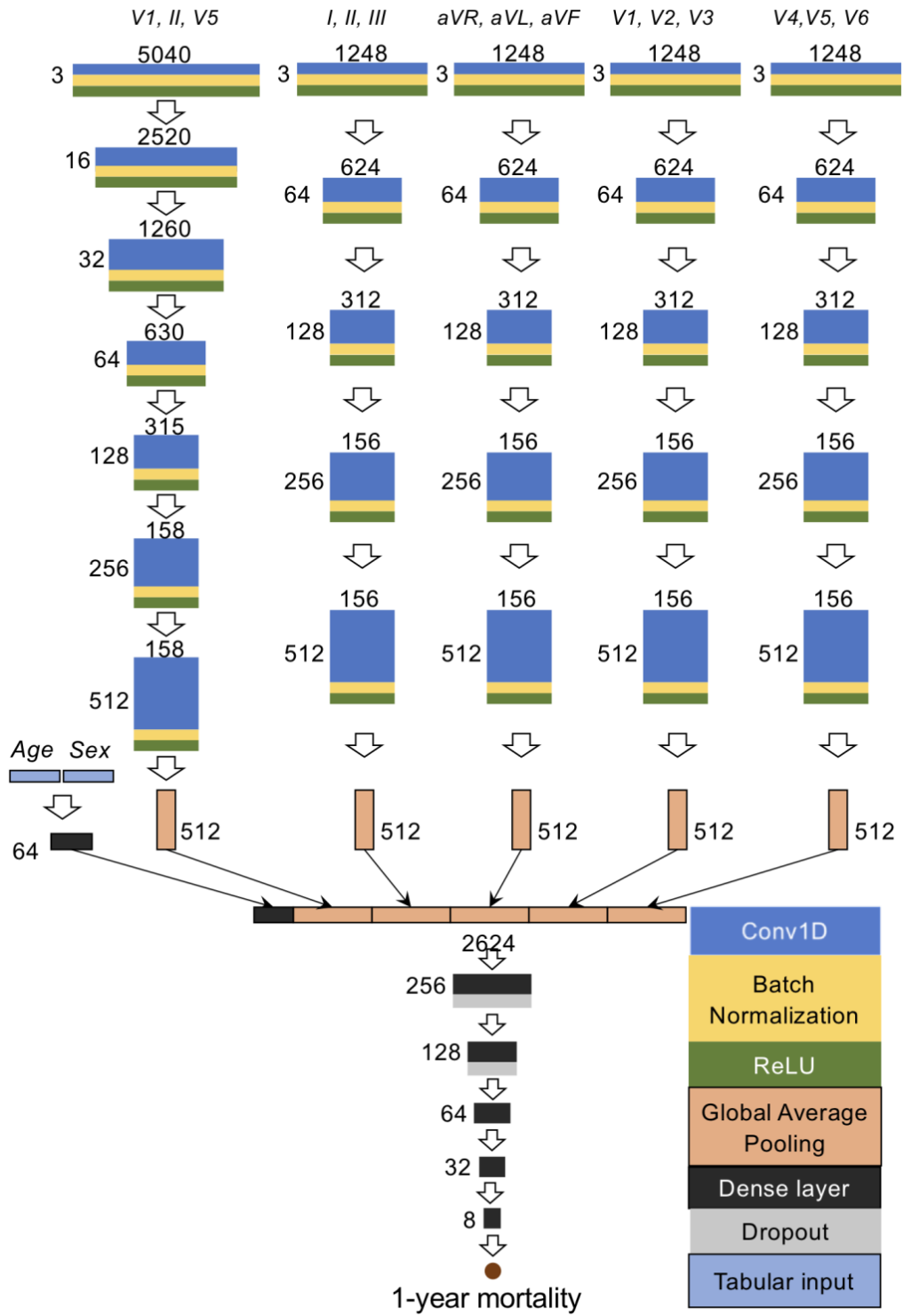

*Figure 4 Model Architecture used in the study.*

Replacing the GAP layer with long short-term memory[26] gave similar performance for significantly longer run times; hence, the final model used GAP layers.

We evaluated loss (binary cross-entropy) on the validation set for each epoch. The training was terminated if the validation loss did not decrease for 10 epochs (early-stopping criteria), and the maximum number of epochs was set to 500. The model was implemented using Keras (version: 2.1.6-tf) with a TensorFlow backend (version: 1.9.0) in python (version: 3.5.2) and default training parameters were used except where specified. For single leads as input, a single branch of the abovementioned model was used. When demographic variables (age and sex) were added to the model, a 64 hidden unit layer following the input layer was concatenated with other branches. All training was performed on an NVIDIA DGX1 platform with eight V100 GPUs and 32 GB of RAM per GPU. We independently fit each fold on 5 GPUs and each epoch took ~10 mins. We evaluated the performance of the model with 5-fold cross-validation to predict 1-year mortality.

To compare the prognostic efficacy of the ECG voltage-time traces with corresponding clinically reported "ECG measures", we cross-validated an XGB classifier with the same folds identified above. In order to fairly compare the performance of XGB models to the DNN that ingested the voltage-time data, in a paired fashion, we re-trained the DNNs using only the studies (approximately 75% of the total) which had ECG measures available. The XGB classifier performed better than a random forest classifier.

*Survival analysis:*

We performed Kaplan-Meier survival analysis[27] with the available follow-up data stratified by the DNN model prediction, using a likelihood threshold of 0.5 as the operating point. The data were censored based on the most recent encounter. We fit a Cox Proportional Hazard model[28] regressing mortality on the DNN model-predicted classification of alive and dead in the subset of normal ECGs and the subset of abnormal ECGs. The mean values and lower and upper bounds of the 95% confidence intervals across the five folds were reported. The *lifelines* package (version: 0.18.6) in python was used for survival analysis.

*Survey*:

In an effort to identify the differential clinical features between true positives and true negatives in ECGs interpreted as normal by a physician, we designed a series of surveys for three independent cardiologists. Pairs of ECGs, including one true positive (correctly predicted by the model to associate with death at 1 year) and one true negative (correctly predicted by the model to associate with survival at 1 year), matched for age and sex, were presented to each cardiologist, blinded to the model outcome. The cardiologists were asked to assess the presented ECGs for the patient at risk of death in a year and were told that one of the two presented ECGs was a true positive and the other a true negative. Sample size required for the 1-sample binomial test for true proportion and null hypothesis proportion equal to 0.6 and 0.5, respectively, with power of 0.8 and type I error rate of 1% was calculated to be 280. With model predictions of likelihood threshold greater than 0.75, 401 true positive ECGs were identified and matched with true negative ECGs with prediction likelihood threshold less than 0.25. The final ECG clinical interpretations (including the 9 computed ECG measurements and findings by the original interpreting cardiologist) were also presented to the cardiologists along with sex and age. After this was completed, the cardiologists were then shown an independent set of 240 pairs of ECGs in the same setup as above with the outcomes shown in order to help them identify potential differential clinical features. Without being told the accuracy or results of the initial survey, each cardiologist then reviewed the original blinded survey again after being shown the survey with marked outcomes.

**Acknowledgements:**

The authors would like to acknowledge Christopher Nevius and Bern McCarty for their help in submitting the IRB for the study and developing a scheduler for efficient computational scheduling of the study experiments.

**Funding sources:**

This work was supported in part by funding from the Pennsylvania Dept of Health, an American Heart Association Competitive Catalyst Award (17CCRG33700289), a grant from the National Institutes of Health (NIH UL1 TR-002550), and the Geisinger Health Plan and Clinic. The content of this article does not reflect the view of the funding sources.

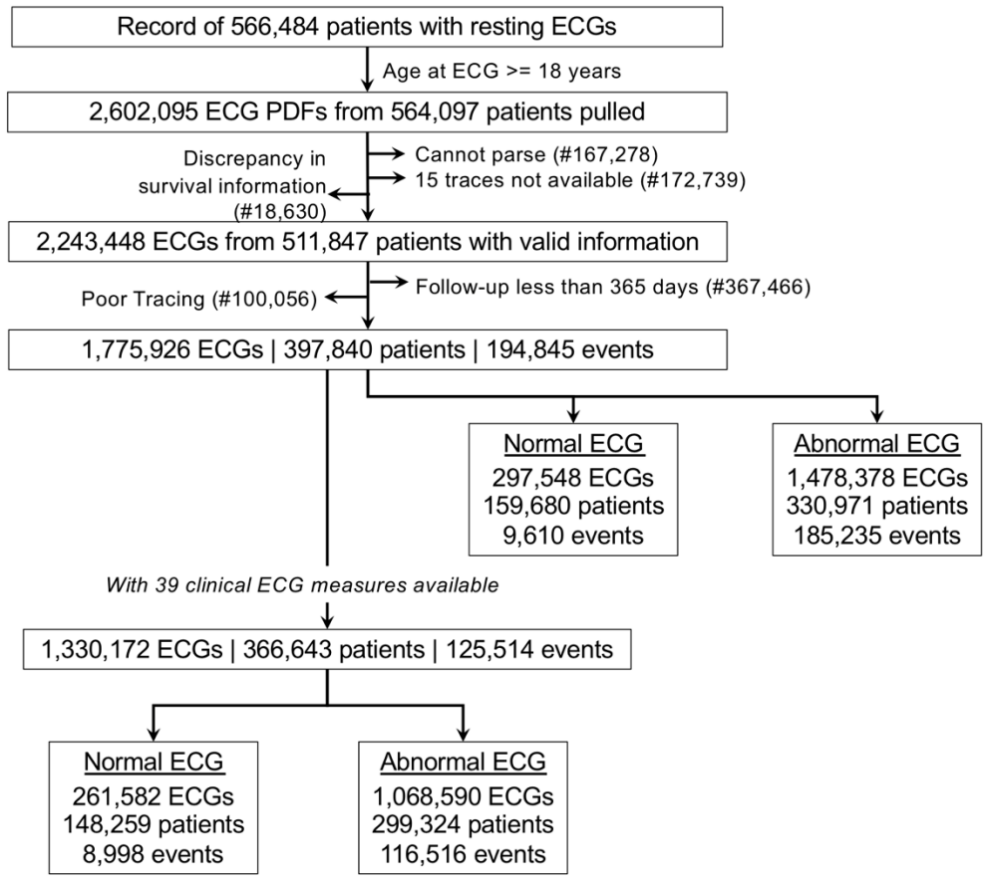

Supplementary Figure 1 Summary of data used in the study.